\documentclass{amsart}

\usepackage{amsmath,amssymb,amsthm,a4wide}
\usepackage{mathtools}
\usepackage{graphicx,tikz}

\theoremstyle{definition}

\theoremstyle{remark}

\begin{document}

\title[]{On the Dual Geometry of\\ Laplacian Eigenfunctions}

\author[]{Alexander Cloninger}
\address{Department of Mathematics, University of California, San Diego, CA 92093, USA}
\email{acloninger@ucsd.edu }

\author[]{Stefan Steinerberger}
\address{Department of Mathematics, Yale University, New Haven, CT 06511, USA}
\email{stefan.steinerberger@yale.edu}

\begin{abstract} We discuss the geometry of Laplacian eigenfunctions $-\Delta \phi = \lambda \phi$ on compact manifolds
$(M,g)$ and combinatorial graphs $G=(V,E)$. The 'dual' geometry of Laplacian eigenfunctions is well understood on
$\mathbb{T}^d$ (identified with $\mathbb{Z}^d$) and $\mathbb{R}^n$ (which is self-dual). The dual geometry is of tremendous
role in various fields of pure and applied mathematics. The purpose of our paper is to point out a notion of similarity
between eigenfunctions that allows to reconstruct that geometry. Our measure of 'similarity' $ \alpha(\phi_{\lambda}, \phi_{\mu})$
between eigenfunctions $\phi_{\lambda}$ and $\phi_{\mu}$ is given by a global average of local correlations
$$  \alpha(\phi_{\lambda}, \phi_{\mu})^2 = \| \phi_{\lambda} \phi_{\mu} \|_{L^2}^{-2}\int_{M}{ \left( \int_{M}{ p(t,x,y)( \phi_{\lambda}(y) - \phi_{\lambda}(x))( \phi_{\mu}(y) - \phi_{\mu}(x)) dy} \right)^2 dx},$$
where $p(t,x,y)$ is the classical heat kernel and $e^{-t \lambda} + e^{-t \mu} = 1$. This notion recovers all classical 
notions of duality but is equally applicable to other (rough) geometries and graphs; many numerical examples in different continuous and discrete settings illustrate the result.
\end{abstract}

\maketitle

\section{Introduction}

\subsection{Introduction.} The Laplacian eigenfunctions on $\mathbb{T}^2$ are easily determined and given by $\phi_k = e^{i\left\langle k, x\right\rangle}$,
where $k \in \mathbb{Z}^2$ is a lattice point. They are orthogonal in $L^2$ and allow for a representation of a function $f \in L^2(\mathbb{T}^2)$.
However, it becomes very quickly apparent that the geometry of $\mathbb{Z}^2$ is not only a convenient enumeration but plays a fairly
fundamental role itself. Examples include
\begin{enumerate}
\item a beautiful inequality of Zygmund \cite{zyg} stating that for any $r > 0$
$$  \sum_{ \|k\| = r}{ |\widehat{f}(k)|^2}  \leq 5^{1/2} \|f\|^2_{L^{4/3}(\mathbb{T}^2)}$$
and, more generally,  restriction phenomena in Harmonic Analysis.
\item the analysis of the nonlinear Schr\"odinger equation (see e.g. \cite{tao})
$$ i u_t + \Delta u = |u|^{p-1}u $$
as well as other general nonlinear dispersive equations,
\item the structure of pseudo-differential operators \cite{coifman},
\item the operations of wavelets and spectral filters on images \cite{coif2}
\item or, a personal example \cite{stein}, an inequality  for $f \in C^1(\mathbb{T}^2)$ with mean value 0
$$ \| \nabla f\|_{L^2(\mathbb{T}^2)} \left\| \partial_x f + \sqrt{2} \partial_y f\right\|_{L^2(\mathbb{T}^2)} \geq c \| \nabla f\|^2_{L^2(\mathbb{T}^2)}.$$
\end{enumerate}

It is clear that the torus $\mathbb{T}^d$ is a special case. The same is true for $\mathbb{R}^d$ whose eigenfunctions (now understood in a distributional sense) satisfy the same additive relationship
$$ e^{i\left\langle m, x\right\rangle} e^{i\left\langle n, x\right\rangle} = e^{i\left\langle m + n, x\right\rangle}.$$
This is one of the reasons why so much mathematical analysis is possible on $\mathbb{T}^d$ and $\mathbb{R}^d$ that cannot be easily generalized. A somewhat
philosophical question, intentionally put vaguely, is 
\begin{quote}
whether or not, for a given manifold $(M,g)$ or Graph $G=(V,E)$, there is additional structure in the eigenfunctions beyond orthogonality and the eigenvalue.
\end{quote}

\subsection{Graph Signal Processing}
This question has turned out to be of increasing importance in modern problems of data science for rather obvious reasons: if we are interested in either selecting
or filtering out certain types of substructures and if those substructures tend to be connected to certain types of eigenvectors or eigenfunctions, then one needs to understand that interplay. 
Conversely, 'proximity' of eigenvectors should be indicative of capturing the same kind of phenomenon. This is easily observed
on $\mathbb{T}^d$ and $\mathbb{R}^d$ where eigenvectors $e^{i\left\langle m, x\right\rangle}, e^{i\left\langle n, x\right\rangle}$ are 'close' if $\|m - n\|$ is small
and this corresponds to having oscillations point in roughly the same direction. The importance of this elementary observation is difficult to overstate; take, for example,
$f \in L^2(\mathbb{R}^2)$ and define $g \in L^2(\mathbb{R}^2)$ via restricting the Fourier transform
$$ \widehat{g}(\xi_1, \xi_2) = \chi_{\xi_1 + \xi_2 \geq 0}(\xi_1, \xi_2) \widehat{f}(\xi_1, \xi_2),$$
then this corresponds to a filter with an obvious geometric interpretation. More precisely, the very notion of a Fourier multiplier (and, correspondingly entire branches of mathematics) is intimately entangled with this underlying geometry of the eigenfunctions of the Laplacian.

\begin{quote}
\textbf{Challenge} (Graph Signal Processing)\textbf{.} Given a Graph $G=(V,E)$, define, if possible, an analogous geometry on its eigenvectors.
\end{quote}

This is an absolutely fundamental problem, we refer to \cite{ankenmann, ank2, bron, co1, co2,
co3, co4, co5, sarah, ham, irion, per, saito, shu, shu2, shu3} for recent examples. Moreover, it is not expected that this is always (or even generically) possible -- even in Euclidean
space, one would expect that eigenfunctions on generic domains do not have any distinguishing features except for their eigenvalue; this vague statement is
made precise in different ways in the study of quantum chaos \cite{haa, nonne}.

\subsection{Our contribution.} The purpose of this paper is to present a somewhat curious definition of a notion of affinity or similarity between two eigenfunctions (or eigenvectors). This notion is identical on manifolds and graphs and results in a number in $[0,1]$ with 1 indicating strong similarity and 0 denoting weak similarity. This then allows us to take a finite set of eigenfunctions $\left\{ \phi_1, \dots, \phi_n \right\}$ and compute
an $n \times n$ matrix $A \in [0,1]^{n \times n}$ where $A_{ij}$ denotes the similarity of $\phi_i$ and $\phi_j$. This is then interpreted as the weighted affinity matrix of a complete weighted Graph on $\left\{ \phi_1, \dots, \phi_n \right\}$ that we hope encodes the underlying geometry of the eigenfunctions. We then use a fairly standard visualization technique that embeds complete weighted graphs into Euclidean space in a geometry-preserving manner; there is some flexibility in this step and one
could use any number of visualization techniques. The main purpose of this paper is to discuss this algorithm in detail and show that it recovers the geometry of the eigenfunctions in all classical cases where such a geometry exists. This has substantial theoretical and practical applications.
\begin{enumerate}
\item Practically, this allows us to group eigenvectors of a Laplacian into various natural geometric substructures
%; put differently, it allows us to identify intrinsic structure of eigenvectors 
beyond just ordering by their eigenvalue.  This is of obvious significance in Graph Signal Processing, as well as in the choice of eigenvectors used to visualize a number of non-linear dimensionality reduction embeddings.
\item Theoretically, it raises a curious connection between the geometry of eigenfunctions, how the pointwise product $\phi_{i} \phi_{j}$ spreads over the spectrum (i.e. what $\left\langle \phi_i \phi_j, \phi_k\right\rangle$ looks like as a sequence in $k$, see also \cite{stein2}) and, through its definition as a local corellation, the local structure. This gives rise to a number of fascinating theoretical prolems.
\end{enumerate}

\section{A Notion of Similarity} 
\subsection{Similarity.} We define a quantitative notion of similarity $0 \leq \alpha(\phi_{\lambda}, \phi_{\mu}) \leq 1$ between two eigenfunctions. We first define it on compact manifolds $(M,g)$ without boundary (this assumption can be dropped but simplifies exposition). We denote
the solution of the heat equation
\begin{align*}
(\partial_t - \Delta) u(t,x) &= 0\\
u(0,x) &= f(x)
\end{align*}
by $e^{t\Delta}f$. This induces a heat kernel $p(t,x,y)$ satisfying
$$ \left[ e^{t\Delta}f \right] (x) = \int_{M}{ p(t,x,y)f(y) dy}.$$
We have $p(t,x,y) = p(t,y,x)$ and conservation of the $L^1-$mass implies that $p(t,x,\cdot)$ is a probability distribution. Moreover,
Varadhan's short-time asymptotic implies that for $t$ small, $p(t,x,\cdot)$ is essentially a Gaussian centered at $x$ and scale $\sim \sqrt{t}$.
We introduce the following measure of similarity between two eigenfunctions
$$  \alpha(\phi_{\lambda}, \phi_{\mu})^2 = \| \phi_{\lambda} \phi_{\mu} \|_{L^2}^{-2}\int_{M}{ \left( \int_{M}{ p(t,x,y)( \phi_{\lambda}(y) - \phi_{\lambda}(x))( \phi_{\mu}(y) - \phi_{\mu}(x)) dy} \right)^2 dx},$$
where $t$ is the unique solution of 
$$e^{-t \lambda} + e^{-t \mu} = 1.$$
This is \textit{not} a metric. The main motivation for this 
quantity is that
\begin{enumerate}
\item it can be interpreted as an average over local correlations: eigenfunctions that behave similar should look locally similar in lots of places and
\item that it appeared somewhat naturally in studies on products of eigenfunctions \cite{stein2} where it was shown to satisfy the identity
$$ \alpha(\phi_{\lambda}, \phi_{\mu}) = \frac{\|e^{t\Delta} (\phi_{\lambda} \phi_{\mu}) \|}{\|\phi_{\lambda} \phi_{\mu} \|_{L^2}}$$
for exactly $e^{-t \lambda} + e^{-t \mu} = 1$.
\end{enumerate}
The second property immediately shows
$$ 0 \leq \alpha( \phi_{\mu}, \phi_{\lambda}) \leq 1.$$

Both local correlations as well as the diffusion under the heat equation (and thus, implicitly, the distribution
in the spectrum) is measured by the same quantity $\alpha$. It being large implies strong local correlations and frequency transport to lower frequencies (meaning that the
expansion of the product $\phi_{\lambda} \phi_{\mu}$ into eigenfunctions contains some low-frequency contributions) and, conversely, it being small implies frequency
transport to higher frequencies (see \cite{stein2} for details). This motivated it as an interesting object worthy of further study and the investigations reported on in this paper.

\subsection{Creating a Landscape.}
We observe that the definition $\alpha(\phi_{\mu}, \phi_{\lambda})$ only gives rise to a weighted graph, however, we would very much like to understand its intrinsic structure.
This leads to an a very substantial problem that is currently receiving a great deal of interest: how to 'accurately' map vertices of a weighted graph to $\mathbb{R}^d$ in such
a way that vertices that are 'close to each other' are also close in the embedding and, conversely, vertices that are far away are will be mapped to different regions in space (see Figure 1).
We often want $d \in \left\{2,3\right\}$ for visualization purposes.
\vspace{-10pt}
\begin{center}
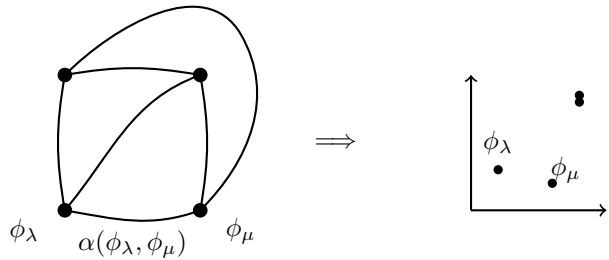
\begin{figure}[h!]
\begin{tikzpicture}[scale =1.8]
\filldraw (0,0) circle (0.05cm);
\filldraw (1,0) circle (0.05cm);
\filldraw (0,1) circle (0.05cm);
\filldraw (1,1) circle (0.05cm);
\draw [thick] (0,0) to[out=350, in = 200] (1,0);
\draw [thick] (0,0) to[out=50, in = 200] (1,1);
\draw [thick] (0,0) to[out=100, in = 260] (0,1);
\draw [thick] (1,0) to[out=80, in = 280] (1,1);
\draw [thick] (1,0) to[out=45, in = 300] (1.3, 1.3) to[out =120, in =45] (0,1);
\draw [thick] (0,1) to[out=10, in = 170] (1,1);
\node at (-0.3, -0.15) {$\phi_{\lambda}$};
\node at (1.3, -0.15) {$\phi_{\mu}$};
\node at (0.5, -0.25) {$\alpha(\phi_{\lambda},\phi_{\mu})$};
\node at (2, 0.5) {$\implies$};
\draw [thick,->] (3,0) -- (3,1);
\draw [thick,->] (3,0) -- (4,0);
\filldraw (3.2,0.3) circle (0.03cm);
\node at (3.2, 0.5) {$\phi_{\lambda}$};
\filldraw (3.6,0.2) circle (0.03cm);
\node at (3.7, 0.3) {$\phi_{\mu}$};
\filldraw (3.8,0.8) circle (0.03cm);
\filldraw (3.8,0.85) circle (0.03cm);
\end{tikzpicture}
\caption{The problem of finding a mapping from the vertices of a weighted graph to $\mathbb{R}^d$ in such a way that 'nearby' vertices get mapped to 'nearby' vertices is of substantial interest
in data science and also occurs in our context.}
\end{figure}
\end{center}

There are a number of methods for creating a low dimensional embedding of points given some notion of distance or similarity between points \cite{belkin2,lafon,kruskal,jolliffe,pearson,saul}. Ultimately,
if the structure is well encoded in the mutual distances, then it does not matter very much which method is used. Throughout this paper, we only use one of the very simplest methods: let us enumerate
the eigenfunctions under consideration by $\left\{1, \dots, n\right\}$ and let 
$$ A_{ij} = \alpha(\phi_i, \phi_j)$$
be the $n \times n$-matrix containing all mutual affinities. We then map
$$ \phi_i \rightarrow\left(v_{1,i}, v_{2, i}, v_{3, i}\right),$$
where $v_1, v_2, v_3$ are the three eigenvectors corresponding to the three largest (in absolute value) eigenvalues of the matrix $A$. The subscript denotes the $i-$th entry of the vector; in particular,
for every eigenfunction $\phi_i$ we use the $i-$th entry in the first, second and third largest eigenvectors as $(x,y,z)-$coordinates for a point in $\mathbb{R}^3$. Sometimes it may be advantageous to not use the first three
eigenvectors but the best result is usually obtained by picking three eigenvectors associated to the lowest eigenvalues. We will also sometimes use
$$ A_{ij} = \alpha(\phi_i, \phi_j)^p \qquad \mbox{for some}~p \geq 1$$
which has the effect of making strong existing affinities even more pronounced by disproportionately  weakening smaller affinities.
 We recall that
$$ \alpha(\phi_i, \phi_j) = \frac{\|e^{t\Delta} (\phi_{i} \phi_{j}) \|_{L^2}}{\|\phi_{i} \phi_{j} \|_{L^2}}$$
requires the computation of a suitable $t$ depending on the eigenvalues of $\phi_i$ and $\phi_j$. We emphasize that the value depends smoothly on $t$ and is not very sensitive. Sometimes
we will even use a fixed value $t_0$, independently of the eigenvalues, for all computations to demonstrate robustness. 
Finally, at points in this section, we also deal with eigenvectors of the normalized graph Laplacian $\mathcal{L}$ defined on a finite collection of points $\left\{x_1, \dots, x_n \right\} \subset \mathbb{R}^d$. This would
correspond to a numerical computation of eigenvectors of a discretization of a continuous manifold. We follow standard procedure and define neighbors of points via the matrix
\begin{align*}
K_{i,j} = \exp\left(-\|x_i-x_j\|_2^2/\sigma^2\right),
\end{align*}
where $\sigma > 0$ is a parameter that fixes a scale.
The normalized graph Laplacian is then denoted
\begin{align*}
L = \mbox{Id}_{n \times n} - D^{-1/2} K D^{-1/2}, & \textnormal{ where } D_{i,j} = \begin{cases} \sum_{j=1}^{n} K_{i,j} , & \textnormal{ for } i=j\\ 0, & \textnormal{ for } i\neq j\end{cases}.
\end{align*}
This normalized Laplacian has an eigendecomposition which satisfies properties similar to those of the manifold Laplacian eigenfunctions; if the points are actually sampled from an underlying manifold, then this construction is known to converge to the continuous Laplacian \cite{belkin1}. We will use this notion of Graph Laplacian throughout the paper when constructing affinities of eigenvectors.

\section{Numerical Examples}
The remainder of the paper is devoted to the study of dual landscapes of Laplacian eigenfunctions and eigenvectors obtained by the method outlined above. More precisely, we will consider
\begin{enumerate}
\item The one-dimensional Torus $\mathbb{T}$ (with endpoints identified) discretized to a cycle graph $C_n$, the eigenfunctions are given by discrete approximations of $\sin{(kx)}$ and $\cos{(kx)}$.
\item Spherical harmonics $\phi_{\ell}^m$ as eigenfunctions of $\mathbb{S}^2$; we recover both indices $\ell$ and $m$.
\item A standard rectangle $[0,4] \times [0,1] \subset \mathbb{R}^2$. Eigenfunctions of the Laplacian are grouped by specifying the number of oscillations in each direction; the method recovers this perfectly.
\item We then study more general cartesian products; the Laplacian eigenfunction of $A \times B$ is merely the product of eigenfunctions on $A$ and eigenfunctions on $B$. The underlying cartesian structure is perfectly recovered even if $A, B$ are rather structure-less objects.
\item On the other end of the spectrum, we consider Erd\H{o}s-Renyi random graphs. Laplacian eigenfunctions on these objects should not exhibit any particular structure nor any distinguishing features. We recover an ordering with respect to the eigenvalue.
\item We conclude by demonstrating several surprising eigenfunctions landscapes, and discuss this method's uses in exploratory spectral graph theory.
\end{enumerate}
We emphasize that, while most examples were computed on the known eigenfunctions, equivalent landscapes are discovered for the empirical eigenvectors of the graph Laplacian on the domain.

%%%%%%%%%%%%%%%%%%%%%%%%%%
 
\subsection{The one-dimensional Torus.}
\begin{figure}[!h]
\includegraphics[height=.45\textwidth]{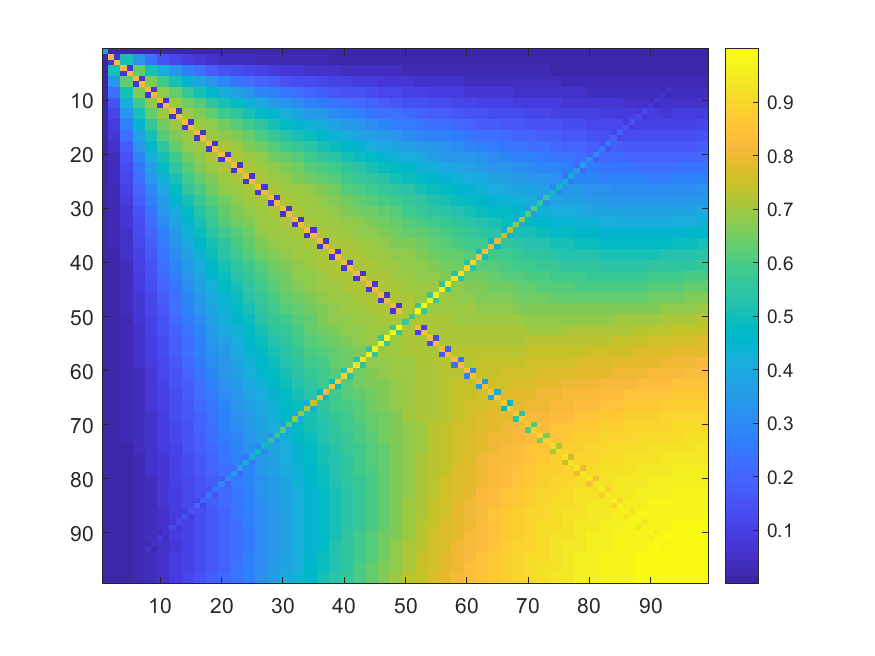}  
\vspace{-20pt}
\caption{Affinity Matrix of the Eigenfunctions on a discretized Torus.}\label{fig:analyticFourier}
\end{figure}

\begin{figure}[!h]
\includegraphics[height=.5\textwidth]{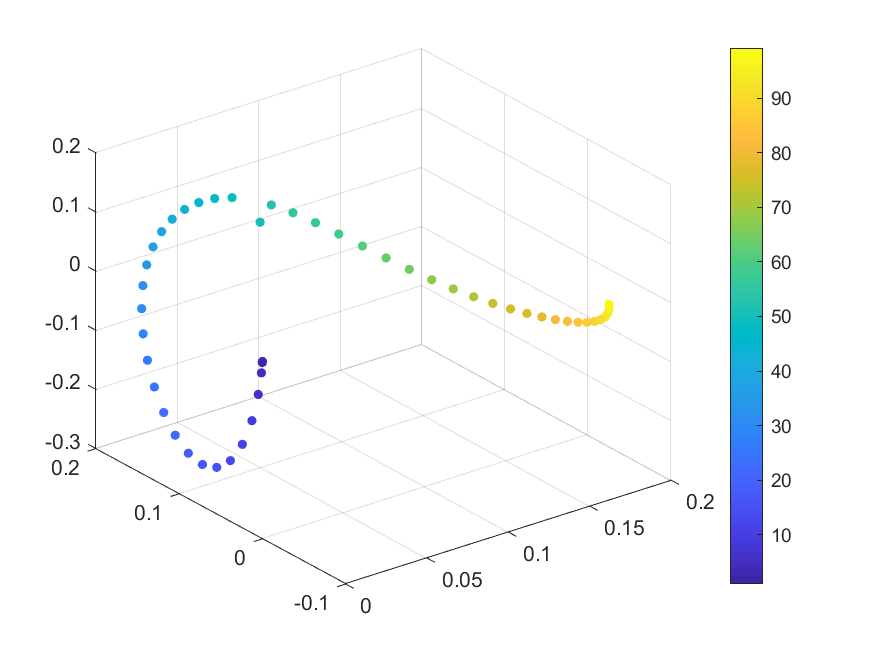} 
\vspace{-20pt}
\caption{Embedding of the distance matrix $A^4$}\label{fig:analyticFourier2}
\end{figure}

We begin with the Laplacian eigenfunctions of the torus restricted to a uniform grid consisting of $n=100$ equispaced points (resulting in $n$ eigenvectors with at least low-frequency eigenvectors approximating the classical trigonometric functions). We emphasize that especially at high frequencies, we only have one sampling point for a wavelength (which is very little). The method nonetheless remains functional (but the effect can be seen in the affinity matrix).
  Figure \ref{fig:analyticFourier} shows the distance matrix $A$, Figure \ref{fig:analyticFourier2} shows the spectral embedding.
\vspace{-10pt}

The method clearly recovers the linear dual structure of the eigenfunctions. It also clearly states that eigenvectors approximating $\sin{kx}$ and $\cos{kx}$ are quite different
functions but orders them in nearby points in the landscape because they behave similarly with respect to other eigenfunctions. Finally, the identity
$$ \exp{\left(i \left[ \frac{n}{2} - k \right] \frac{j}{n} \right)} \exp{\left(i \left[ \frac{n}{2} + k \right] \frac{j}{n} \right)}  = e^{2\pi i j} = 1$$
is automatically discovered. The embedding of the fourth (pointwise) power of the affinity matrix (i.e. $p=4$ as described above) shows a linear structure underlying the eigenvectors.
\vspace{-10pt}

%%%%%%%%%%%%%%%%%%%%%%%%%%

\subsection{Spherical Harmonics}
\begin{figure}[!h]
\includegraphics[height=.45\textwidth]{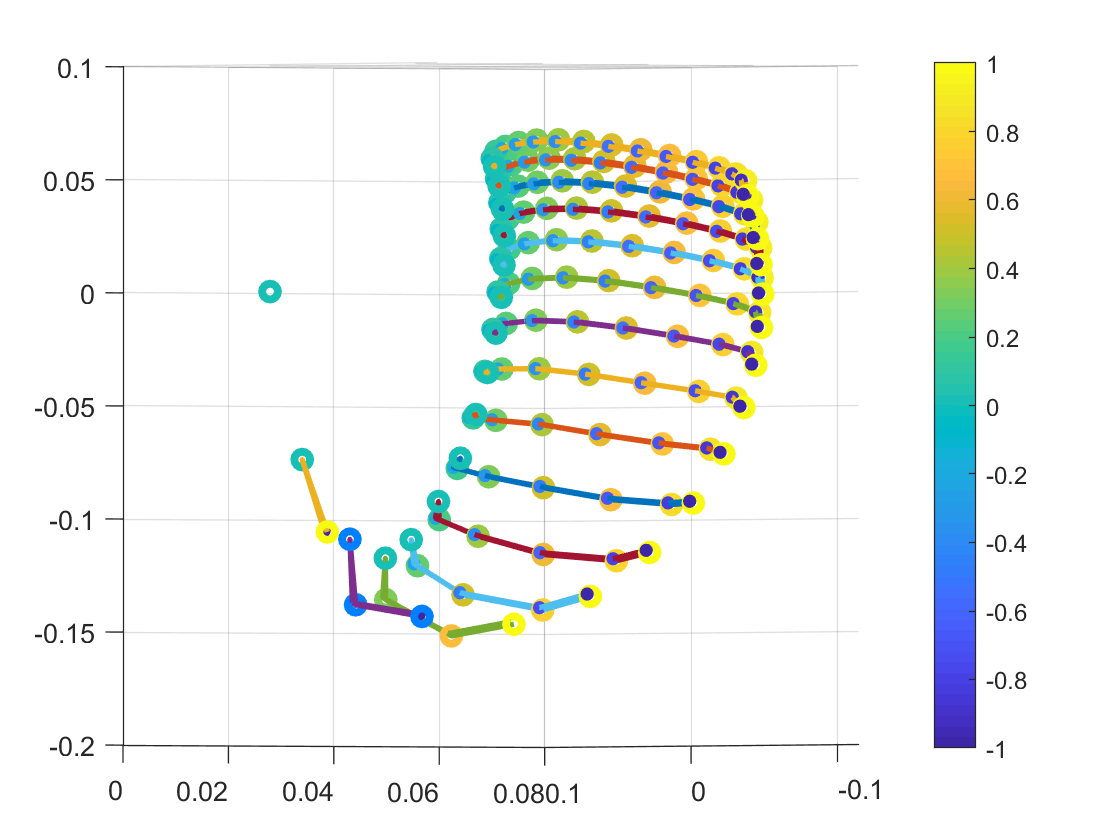} 
%\includegraphics[height=.3\textwidth]{code/AnalyticSphHar/analyticSphHarFixedTime1power_2.png} \\
%$A^{(1)}$ & 
%Embedding of $A^{(2)}$ & 
%Embedding of $A^{(2)}$ 
%Embedding of $A^{(2)}$ with Fixed $t$
\vspace{-13pt}
\caption{Embedding of $A^{(2)}$ for the low frequency spherical harmonics.  For the spectral embedding, each point corresponds to a specific spherical harmonic.  Lines are drawn to connect all $\phi_{\ell}^m$ for a fixed $\ell$, and the color of the node corresponds to $m/\ell$. $\phi_{\ell}^m$ and $\phi_{\ell}^{-m}$ are effectively on top of one another in the embedding, $\phi_{\ell}^m$  for $m<0$ are represented by dots and $\phi_{\ell}^m$ for $m\ge 0$ are circles.}
\label{fig:analyticSphHar}
\end{figure}

We now describe the process on spherical harmonics, i.e. the eigenfunctions of the Laplacian on $\mathbb{S}^2$.  They can be separated into levels $\phi_{\ell}^{m}$, where $-\ell\le m \le \ell$, with corresponding eigenvalue $\ell(\ell+1)$.  We generate these eigenfunctions by taking linear spacing of $181$ points in both spherical coordinate angles $(\alpha, \beta)$, which makes the eigenfunctions satisfy the orthogonality relation
\begin{align*}
\int_0^\pi \int_0^{2\pi} \phi_{\ell}^{m}(\alpha,\beta) \phi_{\ell'}^{m'}(\alpha,\beta) \sin(\alpha) d\beta d\alpha = \delta_{\ell,\ell'} \delta_{m, m'}.
\end{align*}
Figure \ref{fig:analyticSphHar} describes an embedding of the 256 lowest frequency spherical harmonic functions ($\ell \leq 14$).  
\vspace{-23pt}

The method clearly recovers the level $\ell$ of the spherical harmonics.  It also recovers the relative ordering of the degree $m$ of the spherical harmonic, with $\phi_{\ell}^m$ and $\phi_\ell^{-m}$ being found to be quite different functions.  Despite this,  $\phi_{\ell}^m$ and $\phi_\ell^{-m}$ are ordered into nearby points in the landscape as they behave similarly with respect to the rest of the spherical harmonics.

%%%%%%%%%%%%%%%%%%%%%%%%%%%

\subsection{Cartesian Product Structure}

%%%%%%%%%%%%%%%%%%%%%%%%%%%

We consider the rectangle  $[0,4]\times [0,1] \subset \mathbb{R}^2$. The eigenfunctions of the Laplacian with Dirichlet boundary conditions are given by 
$$\phi_{mn} = \sin\left(\frac{mx}{4}\right) \sin\left(\frac{nx}{1}\right) \quad \mbox{ with corresponding eigenvalue} \quad \lambda_{mn} = \frac{m^2}{16} + \frac{n^2}{1^2}.$$
\vspace{-10pt}
\begin{figure}
\begin{center}
\includegraphics[width=0.9\textwidth]{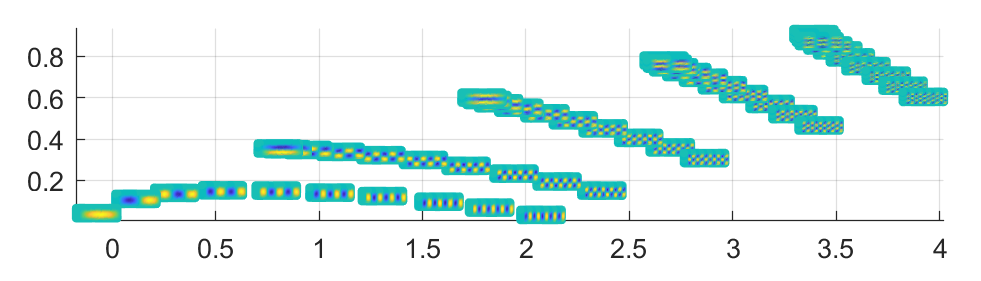}
\end{center}
\vspace{-10pt}
\caption{Low frequency eigenfunctions of the Laplacian on a rectangular grid, with images of each eigenvector displayed in their respective positions. }\label{fig:analyticRectangleImages}
\end{figure}

The precise geometric structure, how many times an eigenfunction oscillates in the $x-$direction vs. how many times it oscillates in the $y-$direction cannot be understood by eigenvalue alone (especially
in the high-frequency limit, these points are equally spaced on an ellipse). Our method clearly recovers the underlying oscillation structure and orders eigenfunctions accordingly.
\begin{figure}[!h]
%\footnotesize
%\begin{tabular}{cc}
%\includegraphics[height=.3\textwidth]{code/AnalyticRectangle/analyticRectangleDistanceMatrixVaryTime.png} &
%\includegraphics[height=.45\textwidth]{code/AnalyticRectangle/analyticRectangleVaryTime1power_5.png} 
\includegraphics[height=.45\textwidth]{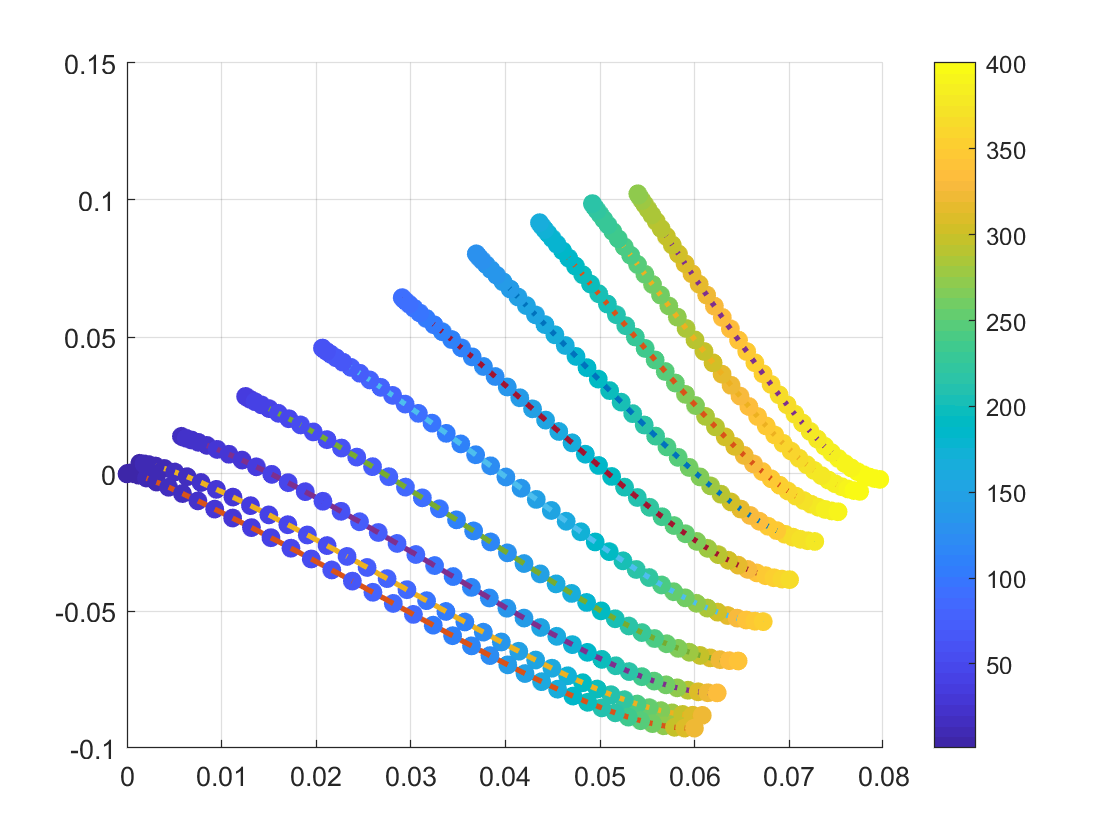} 
%\includegraphics[height=.3\textwidth]{code/AnalyticRectangle/analyticRectangleFixedTime2power_2.png} \\
%%$A^{(1)}$ &  
%Embedding of $A^{(1)}$ & 
%Embedding of $A_{t_0}^{(1)}$ \\
%%\includegraphics[height=.3\textwidth]{code/AnalyticRectangle/analyticRectangleVaryTime1powerWithImages.png} & 
%%\includegraphics[height=.3\textwidth]{code/AnalyticRectangle/analyticRectangleFixedTimeWithImages.png} \\
%%Embedding of $A^{(1)}$ w/ Eig. Image & 
%%Embedding of $A_{t_0}^{(1)}$ w/ Eig. Image  \\
%\end{tabular}
\vspace{-20pt}
\caption{Embedding of eigenfunctions on $[0,4] \times [0,1]$. }\label{fig:analyticRectangle}
\end{figure}

We restricted the exact eigenfunctions to $n=400$ grid points arranged as a $40 \times 10$ rectangle.  The spectral embedding reflects the separable relationship between eigenfunctions oscillating in differing directions, with the $10$ lines of $40$ eigenfunctions perfectly grouping $\phi_{mn}$.   Figure \ref{fig:analyticRectangle} displays the full landscape of the $400$ eigenfunctions. Figure \ref{fig:analyticRectangleImages} also displays the the embedding for $1\le m\le 5$ and $1\le n \le 10$ with each point being the image of the respective $\phi_{mn}$, in order to demonstrate that the spectral embedding does in fact organize the eigenfunctions correctly.

%%%%%%%%%%%%%%%%%%%%%%%%%%
\subsection{More General Cartesian Products}
The reconstruction of tensor product geometry and separable eigenfunctions holds at a much greater level of generality. 
\begin{figure}[!h]
\includegraphics[height=.5\textwidth]{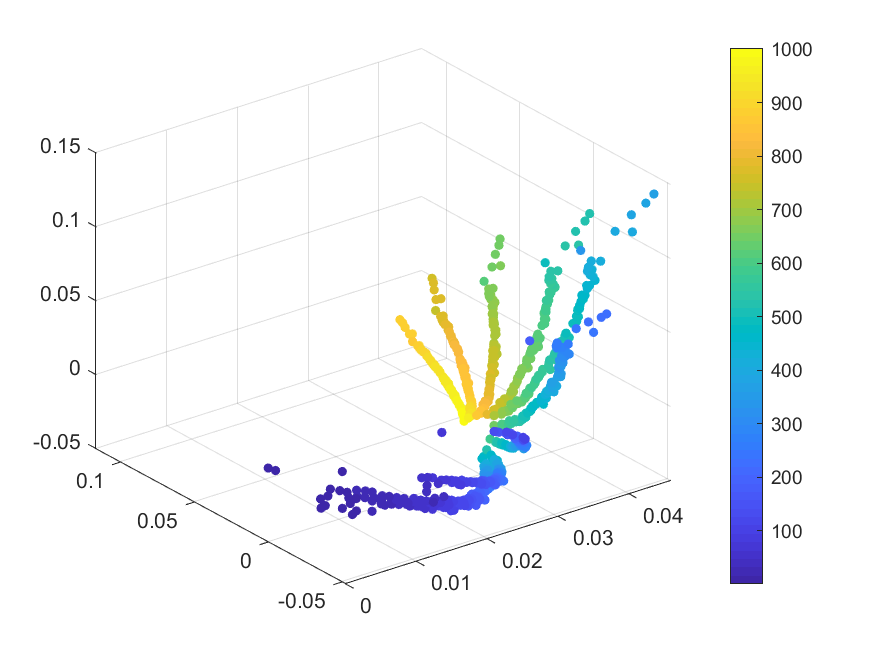} 
\vspace{-20pt}
\caption{Embedding of eigenvectors recovers $X\times Y$ with $x_i\sim \mathcal{N}(0,\sigma^2)$.}\label{fig:empiricalTensorProduct}
\end{figure}
We sample a set $X$ of 100 points in $\mathbb{R}^2$ from a Gaussian distribution $X\sim \mathcal{N}(0,\sigma^2 I_d)$, where $\sigma^2 =0.01$, then take the set $Y$ of 10 equispaced points from $[0,1]$ and consider the set $X \times Y$. The Cartesian Product Structure is perfectly recovered, the result is shown in Figure \ref{fig:empiricalTensorProduct}.
The method clearly recovers the frequency of oscillation in the $Y$ direction, and separates the eigenfunctions into different groups that constructively interfere with one another.  In particular, the point at the tail of the $k^{th}$ line corresponds to the eigenfunction that is constant on $X$ and varies $k$ times in the $Y$ direction.

%%%%%%%%%%%%%%%%%%%%%%%%%%

\subsection{Objects without Structure}
We now consider an Erd\H{o}s-Renyi Graph $G(1000,0.2)$. This is a random object on $n=1000$ vertices with likelihood of two vertices being connected being $p=0.2$. We expect this object to be completely random and eigenfunctions to all behave in a fairly uniform manner. Results of this flavor have been of great interest recently. We refer,
for example, to a rather general result of Rudelson \& Vershynin \cite{rudi} who prove that eigenvectors of graphs with i.i.d. random entries are uniformly flat in the sense of not having any entries substantially larger than $\sim n^{-1/2}$ (up to logarithmic factors). 
\vspace{-10pt}
\begin{figure}[!h]
\includegraphics[height=.45\textwidth]{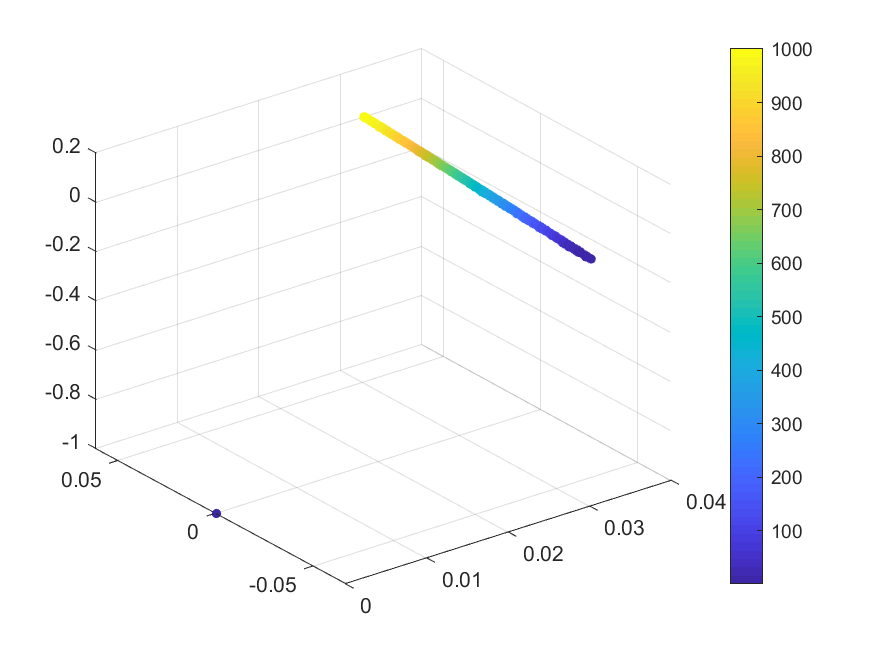} 
\vspace{-20pt}
\caption{Embedding of eigenvectors of the normalized graph Laplacian on Erd\H{o}s-Renyi graph.}\label{fig:empiricalRandomGraph}
\end{figure}

The result of the embedding is shown in Figure \ref{fig:empiricalRandomGraph}. We clearly observe that the first eigenvector (which does not change sign) is separated from the rest but the remaining eigenvectors are fairly structureless and clearly ordered with respect to their eigenvalue.

%%%%%%%%%%%%%%%%%%%%%%%%%%

\begin{figure}[!h]
\begin{tabular}{cc}
\includegraphics[width=.5\textwidth]{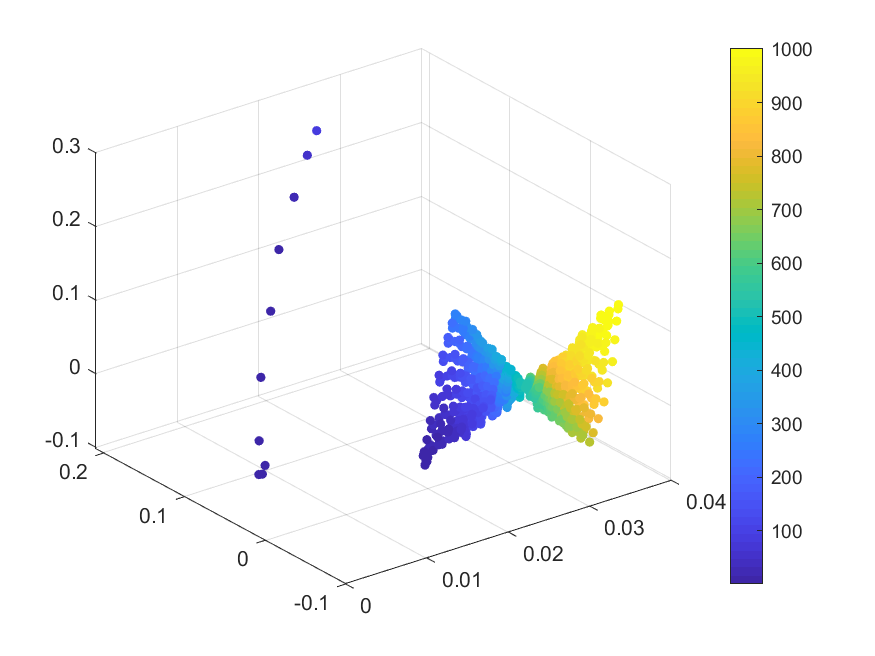} & 
\includegraphics[width=.5\textwidth]{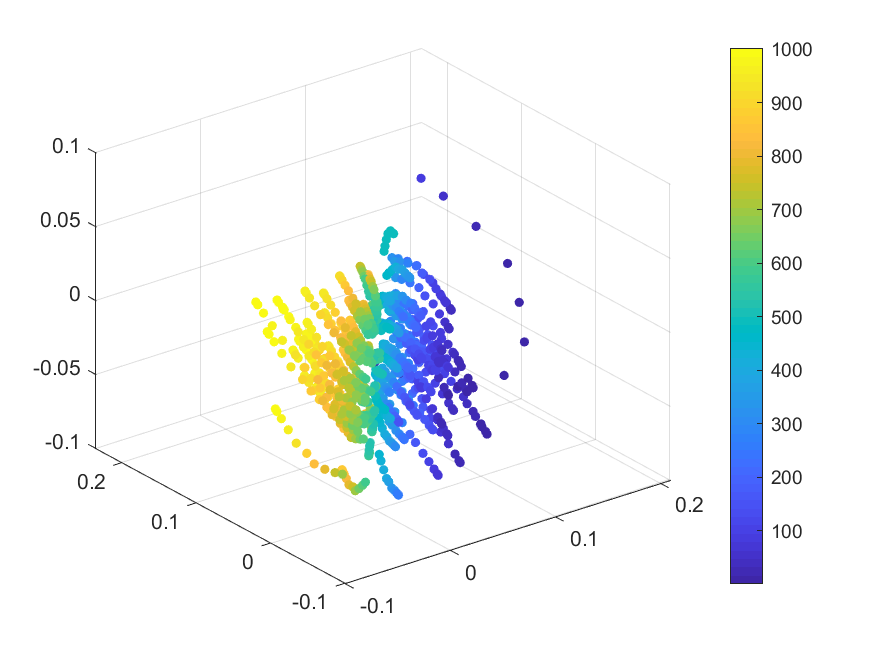}
\end{tabular}
\vspace{-20pt}
\caption{Embedding of eigenvectors recovers $X\times Y$ for cartesian product graph.  (Left) first three coordinates of embedding, (Right) second, third, and forth coordinates of embedding.}\label{fig:empiricalRandomCrossGrid}
\end{figure}

\subsection{Exploratory Spectral Graph Theory}
This technique can be used to discover interesting structures in the eigenspace that are either not obvious or even previously unknown.  We return to the example of separable eigenfunctions: take $K_1$ to be the adjacency matrix of an Erd\H{o}s-Renyi random graph $G(100,0.2)$, and 
$$K_2(x_i, x_j) = e^{-\|x_i - x_j\|^2/\sigma^2}$$
for uniformly sampled points $x_i\in [0,1]$.  We take the final kernel $K$ to be the Kronecker product $K = K_2\otimes K_1$, which corresponds to the Cartesian product of Erd\H{o}s-Renyi grap across $10$ points on a uniformly spaced grid, and build the normalized Laplacian from $K$.   Spectral Theory of cartesian product graphs implies that the eigenfunctions are separable across each dimension \cite{seary}.  Figure \ref{fig:empiricalRandomCrossGrid} shows the landscape of the eigenfunctions as displayed by the three largest eigenvectors of the eigenfunctions affinity matrix, as well as for the second, third, and forth largest eigenvectors. 
The method clearly recovers a very interesting structure to the eigenfunctions, with the lowest frequency eigenfunctions exhibiting a different structure from the majority that are organized in a two-dimensional grid.  Moreover, this landscape can be used to find interesting connections between eigenfunctions that are otherwise non-obvious.  
\vspace{-10pt}
\begin{figure}[!h]
\begin{tabular}{cc}
\includegraphics[width=.5\textwidth]{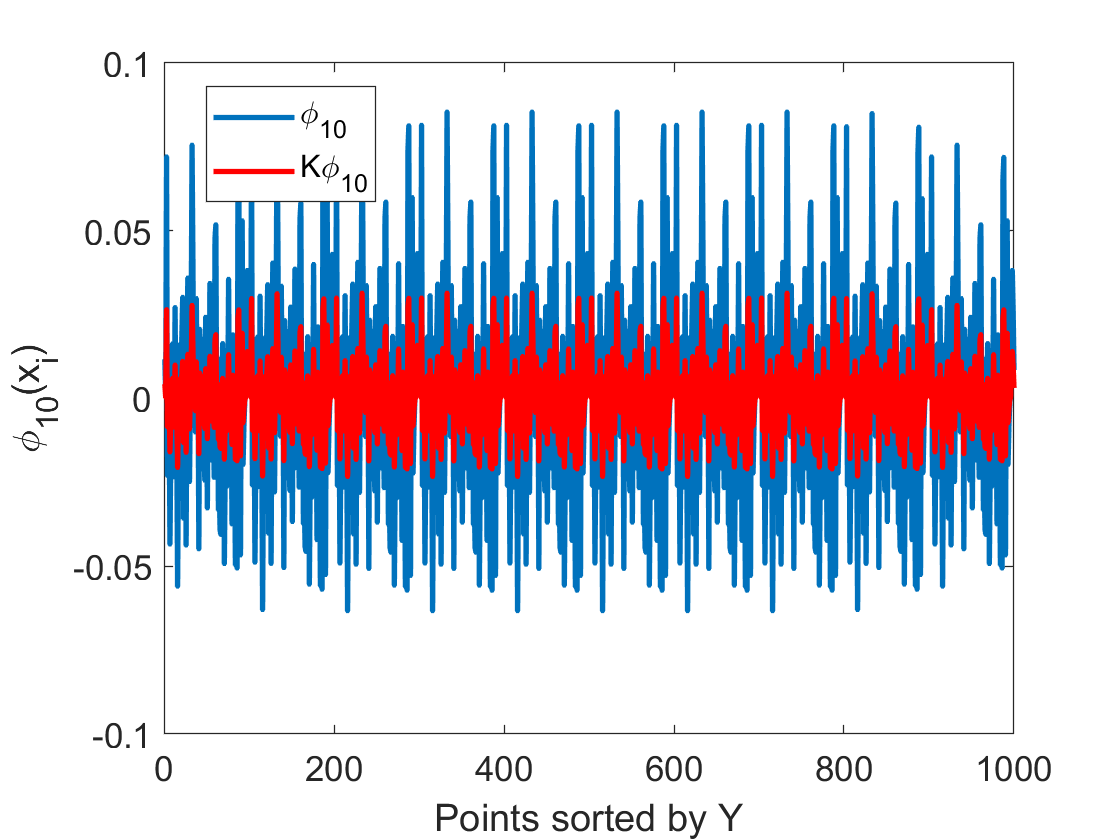} & 
\includegraphics[width=.5\textwidth]{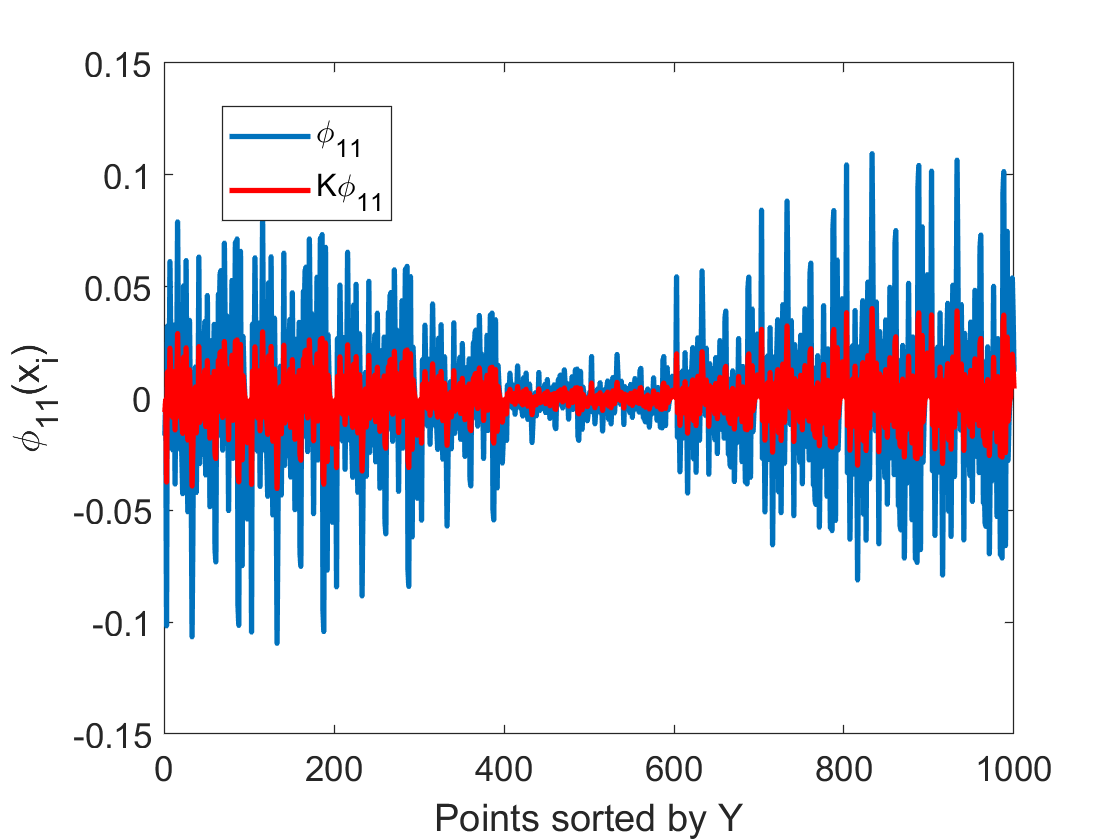} \\
\includegraphics[width=.5\textwidth]{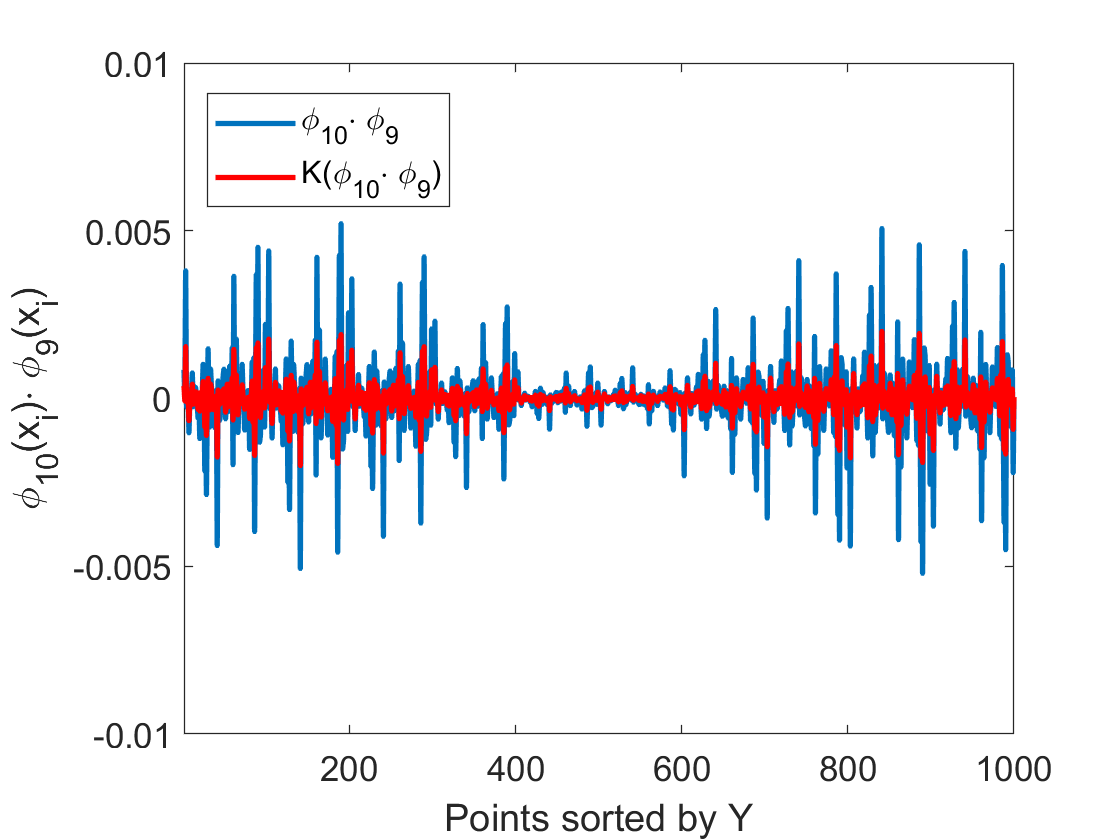} & 
\includegraphics[width=.5\textwidth]{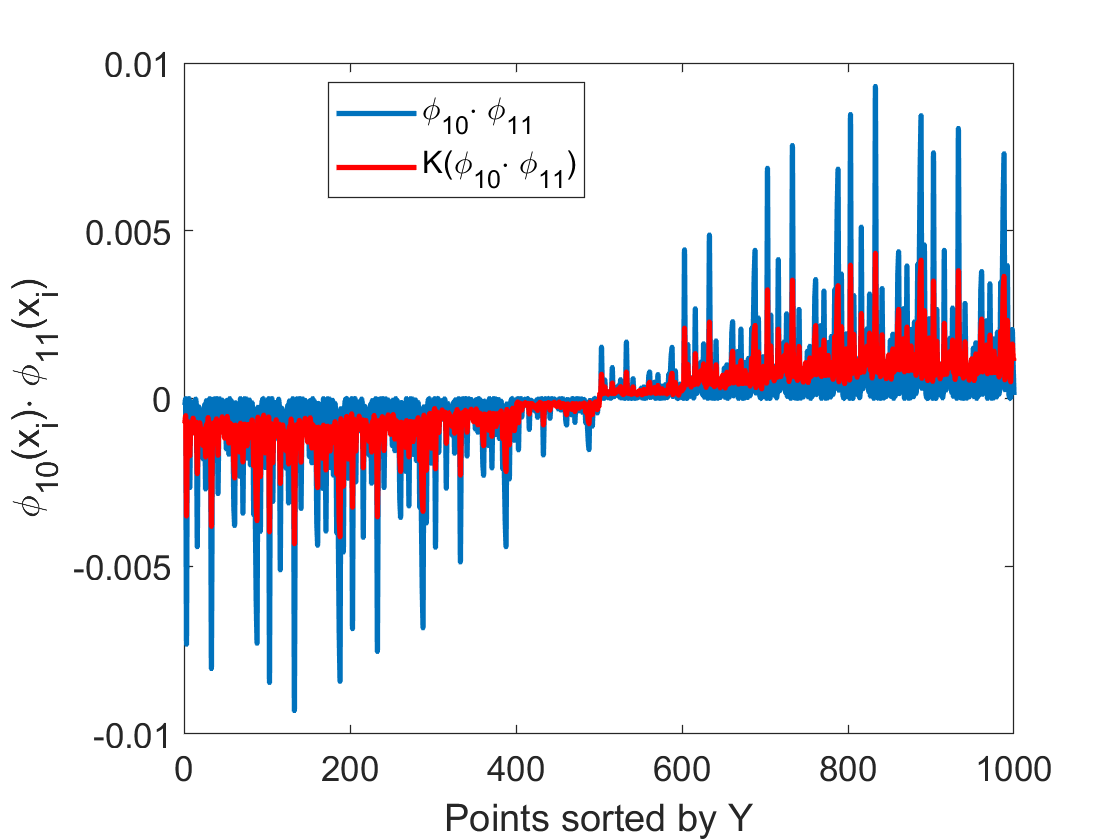} \\
\end{tabular}
\caption{$\phi_{10}$ and $\phi_{11}$ eigenvectors of the cartesian product graph.  In the landscape, the nearest neighbor of $\phi_{10}$ is $\phi_{11}$.  $\phi_9$ is significantly further away in the landscape despite being equally spaced in the spectrum.  }\label{fig:empiricalTensorHadamardExample}
\end{figure}
To demonstrate this, we look at the nearest neighbors of $\phi_{10}$ in the landscape, and examine their Hadamard product with $\phi_{10}$.  $\phi_{11}$ is the nearest neighbor of $\phi_{10}$ in the landscape, and is 75 times closer to $\phi_{10}$ than $\phi_9$ is to $\phi_{10}$ despite the fact that they are about equidistant in the spectrum.  Figure \ref{fig:empiricalTensorHadamardExample} shows these eigenfunctions and their Hadamard product: we discover that despite $\phi_{10}$ and $\phi_{11}$ being chaotic, their product perfect cuts $Y$ in half.  \\

\begin{figure}[!h]
\begin{tabular}{cc}
\includegraphics[width=.5\textwidth]{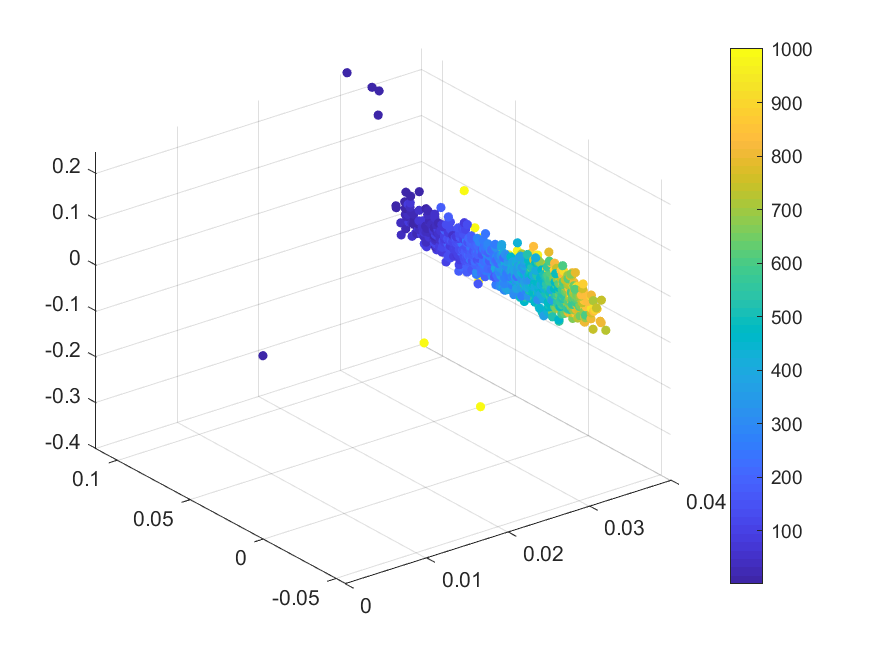} &
\includegraphics[width=.5\textwidth]{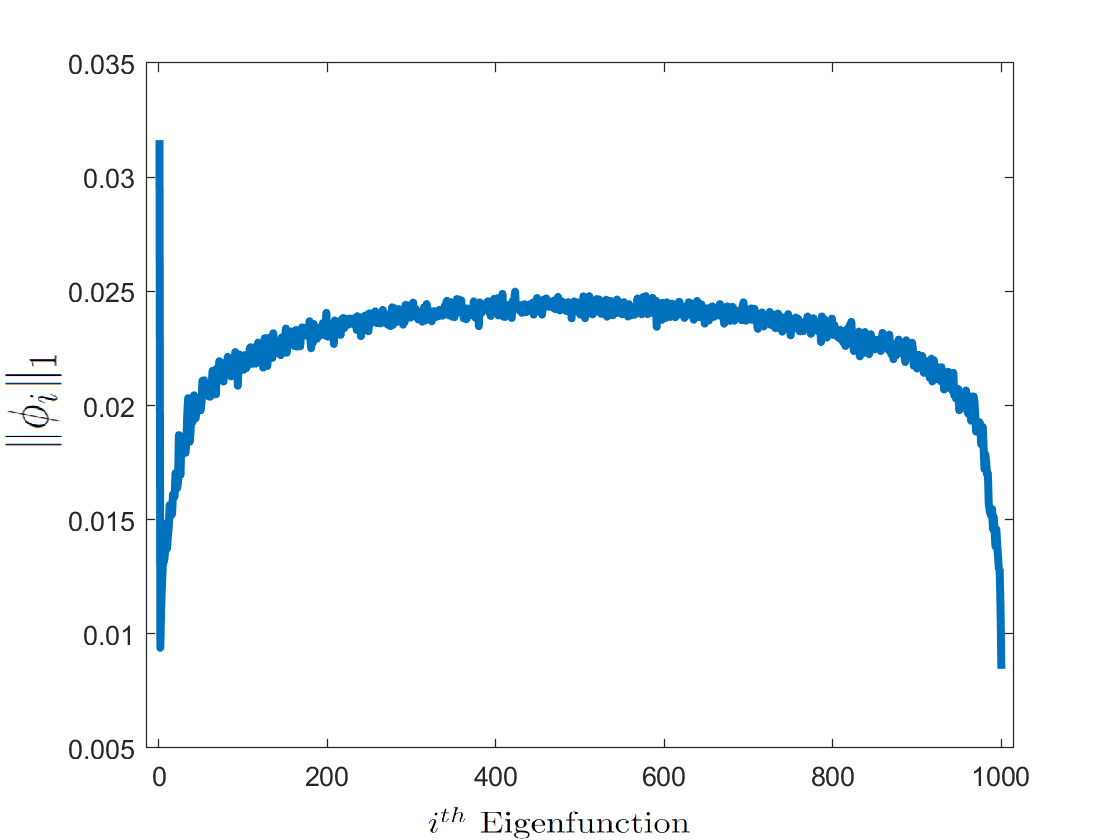} 
\end{tabular}
\caption{Landscape of the eigenfunctions of the unnormalized graph Laplacian on a random graph, and the $\ell^1$ norm of these ($\ell^2-$normalized) eigenfunctions.}\label{fig:unnormalizedErdosLaplacian}
\end{figure}

We also use this technique to look at the differences between the normalized and unnormalized graph Laplacian for and Erd\H{o}s-Renyi random graph $G(1000,0.2)$.  Figure \ref{fig:unnormalizedErdosLaplacian} shows the landscape of the eigenfunctions of the unnormalized graph Laplacian.  Notice that, in contrast to the normalized Laplacian where only $\phi_0$ stands out, in the unnormalized Laplacian there are several  low-freqency and high-frequency eigenfunctions that are clearly separated from the vast majority. 
This was of curiousity to the authors, and upon further investigation, it was discovered that this deviation corresponded to the fact that some eigenfunctions of the unnormalized Laplacian have slightly smaller $\ell^1$ norm than the vast majority. Indeed, as can be seen in Figure \ref{fig:unnormalizedErdosLaplacian}, we observe that the $\ell_1$ seems to have a nontrivial limiting distribution over the spectrum. We are not aware of any results in that direction.

\end{document}